\documentclass[aip, apl,
               amssymb, 
               reprint,
               notitlepage,
               twocolumn,
               a4paper, 
               superscriptaddress]{revtex4-2}

\usepackage[pdftex, dvipsnames]{xcolor}
\usepackage[T1]{fontenc} 
\usepackage{chemformula} 

\usepackage{makecell,tabularx}
\setcellgapes{3pt}
\usepackage{braket}

\usepackage[separate-uncertainty=true, 
            multi-part-units=single,
            ]{siunitx} 

\DeclareSIUnit{\sample}{S}
\DeclareSIUnit\bit{b}

\usepackage{xargs}
\usepackage[disable,colorinlistoftodos,prependcaption,textsize=tiny]{todonotes}
\newcommandx{\unsure}[2][1=]{\todo[linecolor=red,backgroundcolor=red!25,bordercolor=red,#1]{#2}}
\newcommandx{\change}[2][1=]{\todo[linecolor=blue,backgroundcolor=blue!25,bordercolor=blue,#1]{#2}}
\newcommandx{\FIXME}[1][1=]{\todo[linecolor=blue,backgroundcolor=blue!25,bordercolor=blue,#1]{FIXME}}
\newcommandx{\info}[2][1=]{\todo[linecolor=OliveGreen,backgroundcolor=OliveGreen!25,bordercolor=OliveGreen,#1]{#2}}
\newcommandx{\improvement}[2][1=]{\todo[linecolor=Plum,backgroundcolor=Plum!25,bordercolor=Plum,#1]{#2}}
\newcommandx{\addref}[1][1=]{\todo[linecolor=red,backgroundcolor=red!25,bordercolor=red,#1]{Add reference}}

\newcommand\gtwo[2][]{\ifthenelse{\isempty{#2}}{$g^{(2)}(0)$}{\ifthenelse{\isempty{#1}}{$g^{(2)}(0)=#2$}{$g^{(2)}(0)=#2\pm#1$}}}

\begin{document}
\title{Indistinguishable Single Photons from Nanowire Quantum Dots in the Telecom O-Band}

\author{Mohammed K. Alqedra}
\email{alqedra@kth.se}
\affiliation{Quantum Nanophotonics, KTH Royal Institute of Technology, Roslagstullsbacken 21, 10691 Stockholm, Sweden}

\author{Chiao-Tzu Huang}
\affiliation{Research Center for Critical Issues, Academia Sinica, Tainan, Taiwan}
\affiliation{Department of Electrophysics, National Yang Ming Chiao Tung University, Hsinchu 30010, Taiwan}

\author{Wen-Hao Chang}
\affiliation{Research Center for Critical Issues, Academia Sinica, Tainan, Taiwan}
\affiliation{Department of Electrophysics, National Yang Ming Chiao Tung University, Hsinchu 30010, Taiwan}

\author{Sofiane Haffouz}
\affiliation{National Research Council of Canada, Ottawa, Ontario, Canada, K1A 0R6}

\author{Philip J. Poole}
\affiliation{National Research Council of Canada, Ottawa, Ontario, Canada, K1A 0R6}

\author{Dan Dalacu}
\affiliation{National Research Council of Canada, Ottawa, Ontario, Canada, K1A 0R6}
\affiliation{University of Ottawa, Ottawa, Ontario, Canada, K1N 6N5}

\author{Ali W. Elshaari}
\affiliation{Quantum Nanophotonics, KTH Royal Institute of Technology, Roslagstullsbacken 21, 10691 Stockholm, Sweden}

\author{Val Zwiller}
\email{zwiller@kth.se}
\affiliation{Quantum Nanophotonics, KTH Royal Institute of Technology, Roslagstullsbacken 21, 10691 Stockholm, Sweden}

\begin{abstract}
On-demand single-photon sources operating at telecom wavelengths are crucial for quantum communication and photonic quantum technologies. In this work, we demonstrate high-purity, indistinguishable single-photon generation in the telecom O-band from an InAsP/InP nanowire quantum dot. We measured a single-photon purity of $g^2(0)=0.006(3)$ under aboveband excitation. Furthermore, we characterize two-photon interference via Hong–Ou–Mandel measurements and achieve a photon indistinguishability of $94.6\%$ with a temporal postselection of 100 ps time window and $5.58\%$ without temporal postselection. We measure a first-lens source efficiency of $\sim28\%$. These results highlight the potential of nanowire quantum dots as a promising source of telecom single photons for photonic quantum applications, offering deterministic positioning, efficient photon extraction, and scalable production.
\end{abstract}
\maketitle
Quantum communication protocols and photonic-based quantum information processing rely on single photons as fundamental carriers of quantum information \cite{Kimble2008Jun, OBrien2009Dec, Knill2001Jan, Zhong2020Dec}. Consequently, a crucial requirement for realizing these applications is an efficient single-photon source capable of generating photons with high brightness, purity, and indistinguishability. Among various solid-state quantum emitters, semiconductor quantum dots (QDs) have emerged as a promising platform for on-demand single-photon generation due to their deterministic emission and compatibility with photonic integration \cite{Senellart2017Nov, Tomm2021Apr, Ding2016Jan, Mnaymneh2020Feb, Elshaari2020May, Descamps2023Oct}. By carefully controlling the growth conditions and composition of QDs, their emission wavelength can be tuned to meet the specific requirements of various quantum photonic technologies. In particular, emission in the telecom wavelength windows is highly desirable due to the low transmission losses and minimal chromatic dispersion in fiber networks, enabling efficient long-distance quantum communication \cite{Yang2024Jul, Zeuner2021Aug}. Generating photons directly in the telecom range also eliminates the need for frequency conversion, which can introduce inefficiencies and noise\cite{Morrison2023Jun, Zahidy2024Jan}. Telecom-wavelength emission has been demonstrated using InAs/InP and InGaAs/GaAs quantum dots \cite{Miyazawa2005May, Muller2018Feb, Nawrath2019Jul, Zeuner2021Aug}, with significant progress made toward enhancing the indistinguishability of sequentially emitted photons through two-photon interference (TPI) measurements. This has been achieved in self-assembled quantum dots by incorporating QD-mesa structures \cite{Vajner2024Feb}, or coupling the emitters to tapered nanobeam waveguides \cite{Rahaman2024Jul} or circular Bragg grating cavities \cite{Joos2024Jul, Ge2024Feb, Holewa2024Apr}, which enhance photon extraction and shorten the radiative lifetime through the Purcell effect, bringing the emission closer to the Fourier transform.

Recently, site-controlled nanowire quantum dots grown via selective-area vapor–liquid–solid epitaxy have emerged as an alternative platform, offering precise spatial positioning and scalable growth in large uniform arrays \cite{Laferriere2022Apr, Wakileh2024Jan, Alqedra2025Feb}. The nanowire geometry can be engineered for efficient light extraction and directional emission, enhancing photon collection efficiency. Furthermore, these structures enable the scalable integration of quantum dot sources with on-chip photonic circuits, making them highly suitable for quantum photonic applications \cite{Mnaymneh2020Feb, Elshaari2020May, Descamps2023Oct, Chang2023Feb, Descamps2024Oct}.

In this work, we demonstrate the on-demand generation of high-purity single photons in the telecom O-band from an InAsP/InP  nanowire quantum dot, achieving a single-photon purity of 0.006(3) under pulsed above-band excitation. We measured the indistinguishability of consecutively emitted photons through two-photon interference measurements, achieving a Hong–Ou–Mandel (HOM) interference visibility of 94.6$\%$ with a temporal postselection of 100 ps and 5.58$\%$ without postselection. In addition, we estimate a first-lens source efficiency of about 28$\%$. These findings highlight the potential of nanowire quantum dots as a bright, telecom-wavelength single-photon source, offering a promising platform for deployment in quantum communication networks and integrated photonic systems.

\begin{figure*}[htb]
\centering

\includegraphics[width=2\columnwidth]{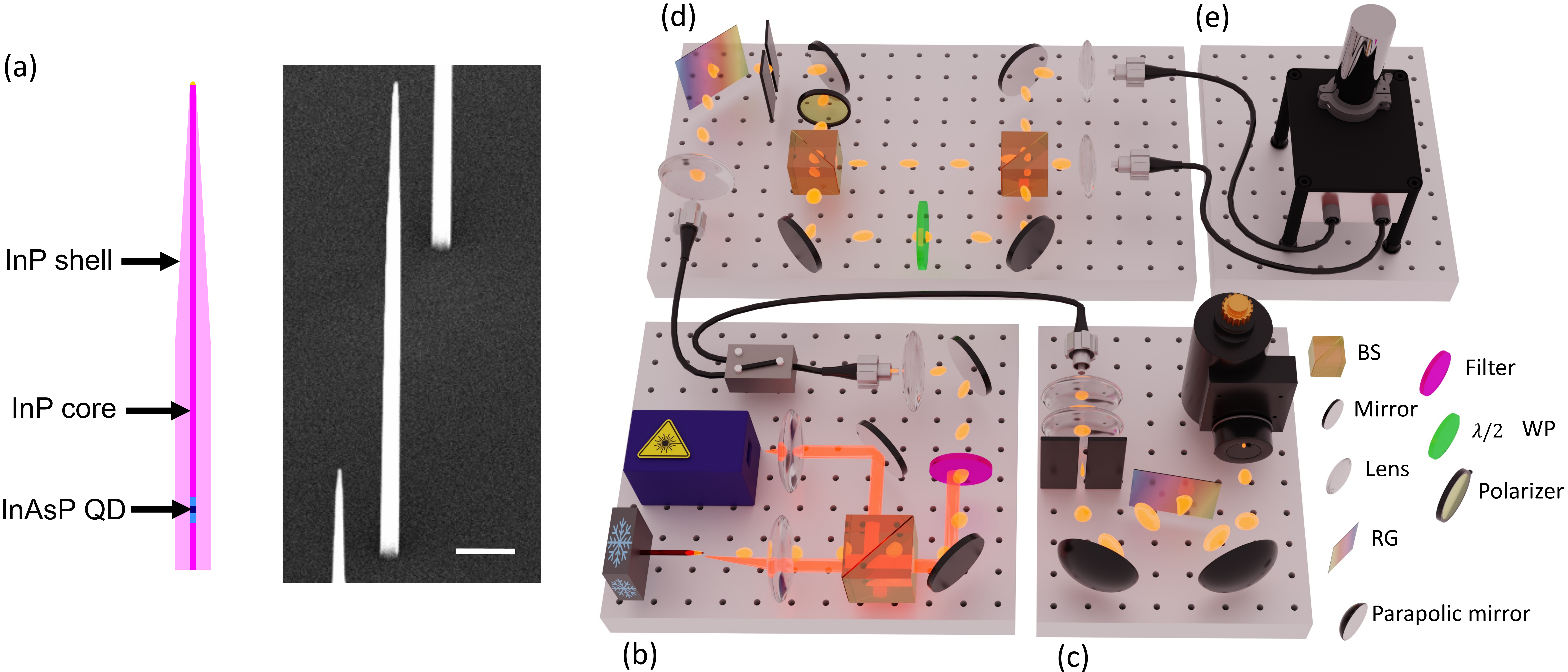}
\caption{\label{fig-setup}
(a) The quantum light source consists of an InP core, an InAsP quantum dot (QD), and an InP shell. The figure shows an SEM image of the nanowire QD on the growth chip. Experimental setup: (b) Confocal microscopy setup, (c) nitrogen-cooled InGaAs spectrometer, (d) HOM setup, and (e) superconducting nanowire single-photon detectors (SNSPDs)}
\end{figure*} 

The InAsP/InP nanowire quantum dot sample investigated in this work was grown using selective-area vapor–liquid–solid (VLS) epitaxy on InP substrate \cite{Dalacu2011Jun}. During the growth, a $\sim3$\,nm thick InAs$_{0.68}$P$_{0.32}$ quantum dot is formed within a $20$\,nm InAs$_{0.5}$P$_{0.5}$ core. The core is subsequently clad with an InP shell to form a photonic waveguide of base diameter $310$\,nm tapered to $20$\,nm over the $12\,\mu$m length of the nanowire \cite{Haffouz2020Sep, Laferriere2023Feb}. This growth method is scalable and enables the production of large, uniform arrays of nanowire quantum dots with similar optical characteristics. Figure \ref{fig-setup} (a) shows a close-up scanning electron microscopy (SEM) image of a nanowire and a schematic showing its structure.

A schematic illustration of the experimental setup is presented in Figure \ref{fig-setup}, showing the confocal microscopy setup (b), nitrogen-cooled InGaAs spectrometer (c), HOM setup (d), and superconducting nanowire single-photon detectors (SNSPDs) (e). The nanowire quantum dot sample was mounted inside a closed-cycle cryostat (Attodry 2100) and cooled down to 1.6 K. An actively power-stabilized 80 MHz tunable picosecond pulsed laser was used to excite the quantum dot. The laser beam was focused onto the QD using an objective lens with a numerical aperture (NA) of 0.8, which  also collected the emitted photons. A 1250 nm long-pass filter was mounted in the collection path to block residual excitation light and pass only the QD emission to be coupled into a single-mode fiber. The QD emission can be routed either to a spectrometer equipped with a liquid nitrogen-cooled InGaAs detector-array for photoluminescence (PL) measurements or to a tunable reflection grating for narrow-band spectral filtering of specific QD excitonic peaks. After filtering, the light was directed to either an autocorrelation setup for second-order correlation measurements or to the HOM interferometer for indistinguishability measurements. For autocorrelation measurements, Hanbury Brown and Twiss (HBT) measurements were performed by splitting the filtered QD emission using a 50:50 fiber beam splitter. The outputs were sent to two superconducting nanowire single-photon detectors (SNSPDs). The arrival times of the detected photons were recorded with a time-to-digital converter (qtools quTAG). Before sending the light to the HOM interferometer, the emitted photons were filtered in polarization using a polarizing beam splitter (PBS), with the transmitted component directed to the interferometer input. In the HOM interferometer, a non-polarizing 50:50 beam splitter (BS) split the photon stream into two paths, introducing a temporal separation of 12.5 ns, corresponding to the laser repetition rate. Light traveling in the two paths was recombined at a second BS, where two photons emitted by the quantum dot from subsequent excitation cycles interfered. The photons at the outputs of the BS were detected by two SNSPDs, and their arrival times recorded using quTAG. Event timing analysis (ETA) \cite{Lin2021Aug} was employed for post-processing, extracting photon correlations, and determining interference visibility.

\begin{figure*}[htb]
\centering

\includegraphics[width=2\columnwidth]{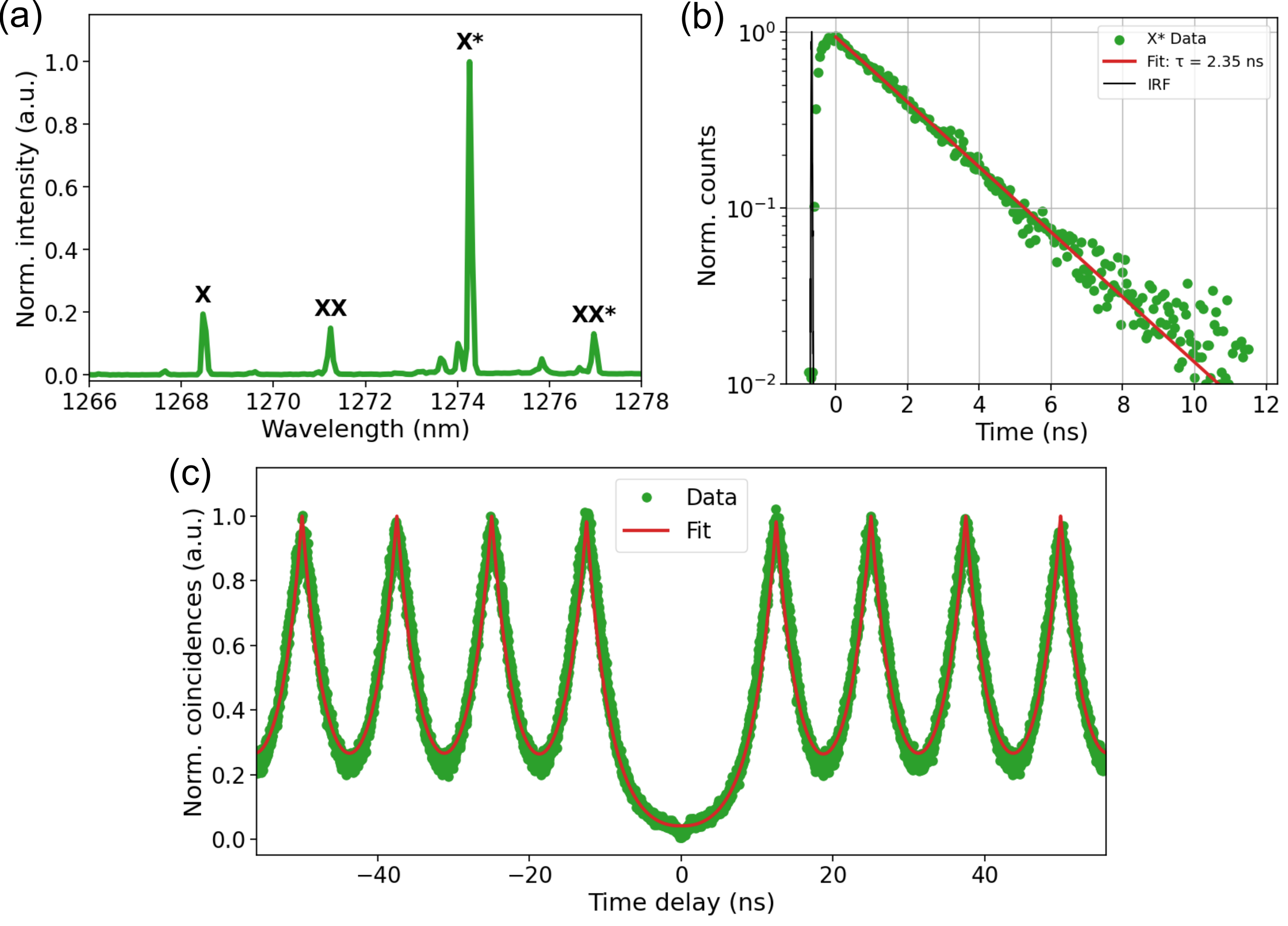}
\caption{\label{fig1-spectra}
\textbf{Spectral characterization of the quantum dot emission } (a) Emission spectra of the nanowire quantum dot under above band excitation at 793 nm. (b) Radiative lifetime of the charged exciton fitted to a single exponential (red line). (c) Autocorrelation function of the charged exciton under above band pulsed excitation at 793 nm wavelength.
}
\end{figure*} 

The PL spectrum of the nanowire quantum dot under aboveband excitation at 793 nm is shown in Figure \ref{fig1-spectra}(a). The labeled peaks correspond to different excitonic transitions, namely the neutral exciton (X), biexciton (XX), charged exciton (X$^*$), and charged biexciton (XX$^*$). The dominance of the charged exciton over other peaks is likely attributed to the dynamics of carrier generation and capture, as detailed in the supporting information in Ref. \cite{Alqedra2025Feb}. In the aboveband excitation scheme, free carriers are generated in the InP host material and subsequently captured by the quantum dot through scattering or phonon-assisted relaxation. Due to differences in electron and hole capture dynamics, an imbalance often occurs, leading to the preferential formation of charged excitons as excess free carriers accumulate in the dot. In contrast, belowband and quasiresonant excitation directly inject carriers into localized states near the quantum dot, minimizing excess free carriers and suppressing the charged exciton contribution. 
We measured a count rate of about 200 kcps on the SNSPDs under 80 MHz pulsed excitation at saturation power. By accounting for the combined collection efficiency of the setup ($\sim1.8\%$) and the SNSPD detection efficiency ($\sim50\%$), we estimate a photon rate of $\sim22\,$MHz at the first lens, corresponding to $28\%$ of the 80 MHz laser repetition rate. This extraction efficiency approaches the upper limit set by the nanowire waveguide geometry, which theoretically allows for up to 
$50\%$ emission in a single direction, demonstrating the effectiveness of the nanowire quantum dot as a highly efficient single-photon source for telecom wavelengths.

The radiative lifetime of the charged exciton was measured by time-resolved photoluminescence at 10$\%$ of the saturation excitation power at 793 nm. The decay was fitted to a single exponential function, which gave a radiative decay time of 2.35(1) ns, as shown in Figure \ref{fig1-spectra}(b). 

\begin{figure*}[htb]
\centering

\includegraphics[width=2\columnwidth]{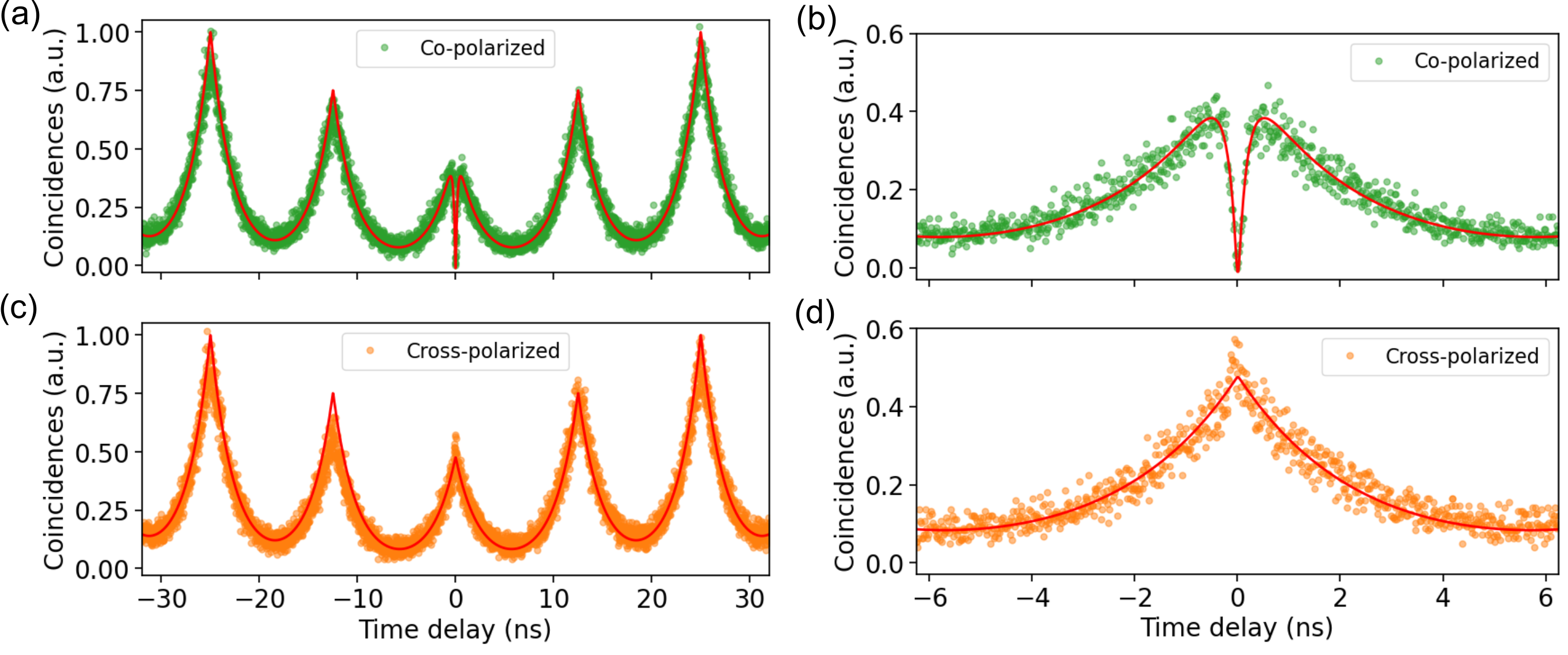}
\caption{\label{fig4-pulsed_visibility_v2}
\textbf{Two photon interference under pulsed excitation at saturation power.} (a) and (b) Co-polarized. (c) and (d) Cross-polarized. (b) and (d) are a zoom into the zero delay peak.}
\end{figure*}

A key figure of merit for single-photon sources is the second-order autocorrelation function at zero time delay, $g^2(0)$, which quantifies the probability of multi-photon emission. An ideal single-photon source should exhibit  $g^2(0)=0$, indicating complete suppression of multi-photon events. To assess the single-photon purity of our nanowire quantum dot, we performed HBT measurement on the charged exciton emission under pulsed excitation at 80 MHz. As shown in Figure \ref{fig1-spectra}(c), the measured autocorrelation function exhibits a periodic structure with a characteristic separation of $T_{rep}=12.5$ ns, corresponding to the laser repetition period. The measured coincidences were fitted to a two-sided exponential model adapted from Ref. \cite{Holewa2024Apr}, omitting the blinking term as its effect is negligible here. The fit function is: 

\begin{equation}
g^{(2)}(\tau) = A.\left(g^{(2)}(0) . e^ { -\frac{|\tau|}{\tau_1} } + \sum_{n \neq 0} e^ { -\frac{|\tau - nT_{rep}|}{\tau_1} }\right),
\end{equation}

where $g^{(2)}(\tau)$ is the coincidences as a function of time delay, A is a normalization factor, $g^{(2)}(0)$ is the zero delay coincidences, $\tau_1$ is the radiative lifetime. To account for the timing resolution of the detection, the model was convolved with a Gaussian instrument response function (IRF) with a full width at half maximum (FWHM) of 50 ps. From the fit we obtain $g^{(2)}(0)=0.006(3)$, indicating a high suppression of multi-photon events and confirming the high purity of the nanowire QD source. The extracted radiative lifetime from the fit is 3.11, which is slightly longer than the independently measured radiative decay of 2.35 ns, obtained from time-resolved PL measurement. This discrepancy is likely due to re-excitation effects at saturation power, $P_{sat}$, during the $g^{(2)}(\tau)$ measurement, whereas the lifetime was measured at $0.1P_{sat}$.

Having established the high purity of the charged exciton, we now investigate the indistinguishability of consecutively emitted photons using HOM interference measurements. The measurements were performed at saturation power under above-band pulsed excitation at 793 nm, where the charged exciton emission is most prominent due to the efficient carrier injection into the quantum dot. Although above-band excitation is generally associated with increased charge noise and spectral broadening, nanowire quantum dots have been shown to mitigate several of these broadening mechanisms due to their structural properties \cite{Laferriere2023Apr}. In particular, the presence of a single emitter per nanowire, the absence of a 2D wetting layer, and the sidewalls formed by epitaxial growth rather than dry etching all contribute to reducing dephasing mechanisms, minimizing charge noise, and improving spectral stability \cite{Laferriere2023Apr, Laferriere2022Apr, Lobl2019Aug, Urbaszek2004Jan}. Linewidths approaching the Fourier transform limit have been achieved in nanowire quantum dots under aboveband excitation, demonstrating that this excitation scheme can still produce highly coherent and indistinguishable single photons \cite{Laferriere2023Apr}.

\begin{figure*}[htb]
\centering

\includegraphics[width=1.5\columnwidth]{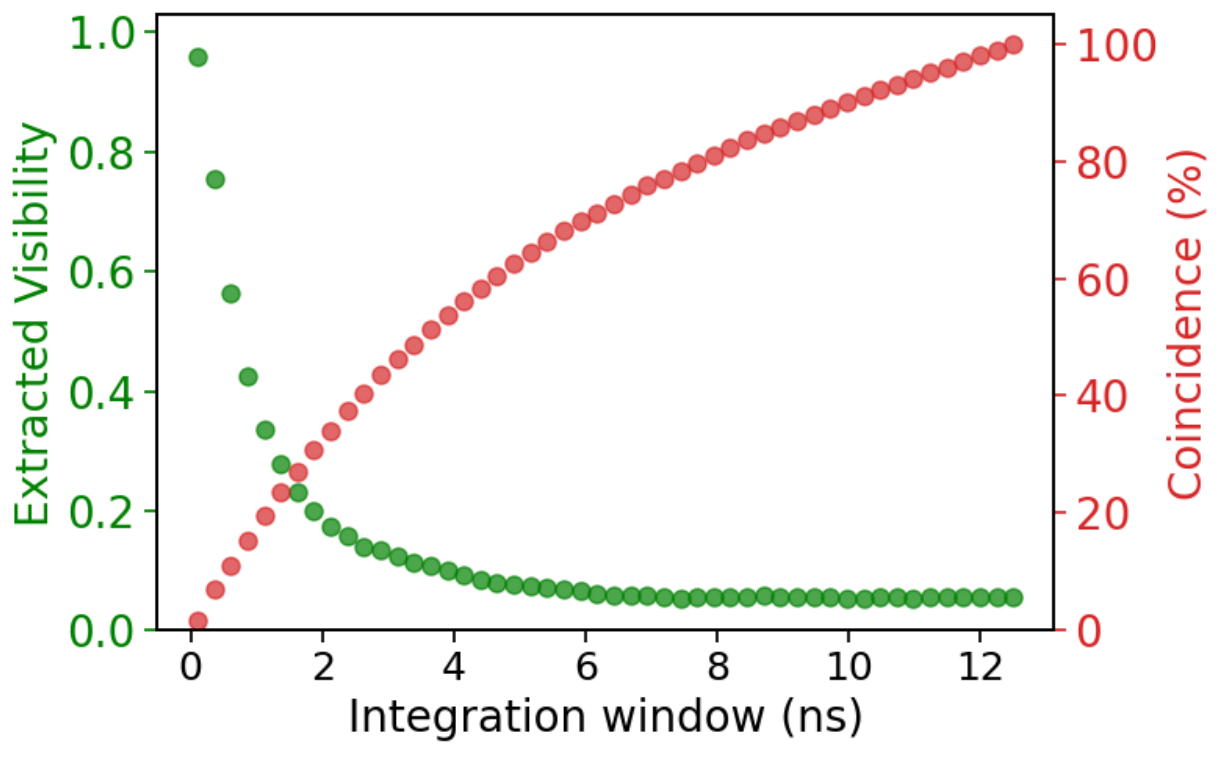}
\caption{\label{fig4-fig5_vis}
Extracted visibility and fraction of coincidences for different integration time windows.
}
\end{figure*}

HOM measurements were performed for both co-polarized and cross-polarized configurations, with polarization control achieved using a half-waveplate in the longer optical path of the interferometer.
Figure \ref{fig4-pulsed_visibility_v2} shows the measured coincidences for co-polarized and cross-polarized configurations. In a pulsed HOM interference experiment, the relative heights of the five central peaks follow a characteristic pattern determined by the degree of photon indistinguishability. For fully distinguishable photons, the zero-delay peak is expected to have half the height of the side peaks at $\pm 2T_{rep}$, while the peaks at $\pm T_{rep}$ are reduced by 25$\%$. In contrast, for fully indistinguishable photons, the zero-delay peak would be fully suppressed, as interfering photon pairs always exit the same output port of the second beam splitter, leading to a suppression of coincidence events. 
Figure \ref{fig4-pulsed_visibility_v2}(a), co-polarized, and (c), cross-polarized, display the five central peaks of the correlation function, with their relative heights reflecting the expected weight distribution. In the co-polarized configuration, the coincidences at zero delay were partially reduced, reflecting partial two-photon interference. Figure \ref{fig4-pulsed_visibility_v2}(b) provides a closer look at the zero-delay region for the co-polarized configuration, revealing a HOM dip that confirms non-negligible photon indistinguishability. The narrow dip is attributed to the finite coherence of the emitted photons, indicating that the system is not at the Fourier transform limit. The high timing resolution of the SNSPDs allows for precise detection of the HOM dip, enabling an accurate characterization of the TPI visibility. We adapted the approach detailed in the supplementary material of Ref. \cite{Vajner2024Feb} to model the co-polarized HOM correlation function without including the blinking term, which we found to be negligible for the investigated nanowire quantum dot under the present experimental conditions. The resulting expression was then convolved with a Gaussian IRF with a 50 ps full width at half maximum to account for the SNSPD timing resolution. From the fit, we extract a time constant of 0.19(9) ns for the HOM dip, which represents the temporal coherence.
The TPI visibility, $V$, was extracted from the raw coincidence histogram by comparing the integrated coincidences at zero delay for co-polarized, $C_{co}(0)$, and cross-polarized, $C_{cross}(0)$, configurations. Specifically, we use:

\begin{equation}
    V \;=\; 1 \;-\; \frac{C_{\mathrm{co}}(0)}{C_{\mathrm{cross}}(0)}\,,
\end{equation}

Without applying any temporal postselection, we obtain a TPI visibility of $5.58\%$ when integrating over the 12.5 ns laser repetition period. By restricting coincidence detection to a narrower time window of 100 ps, we obtain a visibility of $94.6\%$, although the coincidences are significantly reduced for such a narrow time window. To further improve TPI visibility without relying on temporal postselection, resonant or quasi-resonant excitation can be utilized to minimize charge and phonon-induced dephasing from excess carriers in the host material \cite{Gao2024Sep}. Charge stabilization via applied electric fields can further suppress spectral diffusion and charge noise, enhancing photon coherence. Another approach is to shorten the exciton radiative lifetime through Purcell enhancement using a high-Q optical cavity or coupling to a photonic nanostructure, bringing the emission closer to the Fourier transform limit.

Figure \ref{fig4-fig5_vis} shows the TPI visibility (left y-axis) and the fraction of retained coincidence events (right y-axis) as functions of the integration time window. Narrowing the integration window around zero delay progressively excludes more coincidence detections outside the HOM interference region, consequently increasing the extracted TPI visibility. However, this temporal postselection also decreases the percentage of retained events, reflecting a trade-off between maximizing visibility and preserving total signal counts.

We have demonstrated on-demand generation of single photons in the telecom O-band from an InAsP/InP nanowire quantum dot excited above the band gap. Second-order autocorrelation measurements revealed a single-photon purity of 0.006(3). Using HOM interference experiments, we obtain a two-photon interference visibility of 5.58$\%$ without temporal postselection, and 94.6$\%$ when applying a strong temporal postselection. We estimated a first lens source efficiency of 28$\%$. 

These results highlight the potential of site-controlled nanowire quantum dots as single-photon sources for quantum communication, with telecom O-band emission enabling long-distance fiber transmission. The deterministic placement, high extraction efficiency, and scalable production of the nanowires facilitate their integration in photonic circuits, making them suitable sources for quantum photonic applications. While the current temporal coherence is limited, resonant and charge stabilization schemes could improve two-photon indistinguishability\cite{gao2024demand}. Additionally, Purcell enhancement could be utilized to shorten the radiative lifetime, bringing the emission closer to the Fourier transform limit.

\section*{Acknowledgments}
This research was supported by Sweden’s Innovation Agency VINNOVA (Grant No. 2024-00515), and by the National Sciences and Engineering Research Council of Canada (NSERC).

\section*{Data Availability}
The data that support the findings of this study are available from the corresponding authors upon reasonable request.

\bibliography{met_sync}

\begin{thebibliography}{36}%
\makeatletter
\providecommand \@ifxundefined [1]{%
 \@ifx{#1\undefined}
}%
\providecommand \@ifnum [1]{%
 \ifnum #1\expandafter \@firstoftwo
 \else \expandafter \@secondoftwo
 \fi
}%
\providecommand \@ifx [1]{%
 \ifx #1\expandafter \@firstoftwo
 \else \expandafter \@secondoftwo
 \fi
}%
\providecommand \natexlab [1]{#1}%
\providecommand \enquote  [1]{``#1''}%
\providecommand \bibnamefont  [1]{#1}%
\providecommand \bibfnamefont [1]{#1}%
\providecommand \citenamefont [1]{#1}%
\providecommand \href@noop [0]{\@secondoftwo}%
\providecommand \href [0]{\begingroup \@sanitize@url \@href}%
\providecommand \@href[1]{\@@startlink{#1}\@@href}%
\providecommand \@@href[1]{\endgroup#1\@@endlink}%
\providecommand \@sanitize@url [0]{\catcode `\\12\catcode `\$12\catcode `\&12\catcode `\#12\catcode `\^12\catcode `\_12\catcode `\%12\relax}%
\providecommand \@@startlink[1]{}%
\providecommand \@@endlink[0]{}%
\providecommand \url  [0]{\begingroup\@sanitize@url \@url }%
\providecommand \@url [1]{\endgroup\@href {#1}{\urlprefix }}%
\providecommand \urlprefix  [0]{URL }%
\providecommand \Eprint [0]{\href }%
\providecommand \doibase [0]{https://doi.org/}%
\providecommand \selectlanguage [0]{\@gobble}%
\providecommand \bibinfo  [0]{\@secondoftwo}%
\providecommand \bibfield  [0]{\@secondoftwo}%
\providecommand \translation [1]{[#1]}%
\providecommand \BibitemOpen [0]{}%
\providecommand \bibitemStop [0]{}%
\providecommand \bibitemNoStop [0]{.\EOS\space}%
\providecommand \EOS [0]{\spacefactor3000\relax}%
\providecommand \BibitemShut  [1]{\csname bibitem#1\endcsname}%
\let\auto@bib@innerbib\@empty
\bibitem [{\citenamefont {Kimble}(2008)}]{Kimble2008Jun}%
  \BibitemOpen
  \bibfield  {author} {\bibinfo {author} {\bibfnamefont {H.~J.}\ \bibnamefont {Kimble}},\ }\href {https://doi.org/10.1038/nature07127} {\bibfield  {journal} {\bibinfo  {journal} {Nature}\ }\textbf {\bibinfo {volume} {453}},\ \bibinfo {pages} {1023} (\bibinfo {year} {2008})}\BibitemShut {NoStop}%
\bibitem [{\citenamefont {O'Brien}\ \emph {et~al.}(2009)\citenamefont {O'Brien}, \citenamefont {Furusawa},\ and\ \citenamefont {Vu{\ifmmode\check{c}\else\v{c}\fi}kovi{\ifmmode\acute{c}\else\'{c}\fi}}}]{OBrien2009Dec}%
  \BibitemOpen
  \bibfield  {author} {\bibinfo {author} {\bibfnamefont {J.~L.}\ \bibnamefont {O'Brien}}, \bibinfo {author} {\bibfnamefont {A.}~\bibnamefont {Furusawa}},\ and\ \bibinfo {author} {\bibfnamefont {J.}~\bibnamefont {Vu{\ifmmode\check{c}\else\v{c}\fi}kovi{\ifmmode\acute{c}\else\'{c}\fi}}},\ }\href {https://doi.org/10.1038/nphoton.2009.229} {\bibfield  {journal} {\bibinfo  {journal} {Nat. Photonics}\ }\textbf {\bibinfo {volume} {3}},\ \bibinfo {pages} {687} (\bibinfo {year} {2009})}\BibitemShut {NoStop}%
\bibitem [{\citenamefont {Knill}\ \emph {et~al.}(2001)\citenamefont {Knill}, \citenamefont {Laflamme},\ and\ \citenamefont {Milburn}}]{Knill2001Jan}%
  \BibitemOpen
  \bibfield  {author} {\bibinfo {author} {\bibfnamefont {E.}~\bibnamefont {Knill}}, \bibinfo {author} {\bibfnamefont {R.}~\bibnamefont {Laflamme}},\ and\ \bibinfo {author} {\bibfnamefont {G.~J.}\ \bibnamefont {Milburn}},\ }\href {https://doi.org/10.1038/35051009} {\bibfield  {journal} {\bibinfo  {journal} {Nature}\ }\textbf {\bibinfo {volume} {409}},\ \bibinfo {pages} {46} (\bibinfo {year} {2001})}\BibitemShut {NoStop}%
\bibitem [{\citenamefont {Zhong}\ \emph {et~al.}(2020)\citenamefont {Zhong}, \citenamefont {Wang}, \citenamefont {Deng}, \citenamefont {Chen}, \citenamefont {Peng}, \citenamefont {Luo}, \citenamefont {Qin}, \citenamefont {Wu}, \citenamefont {Ding}, \citenamefont {Hu}, \citenamefont {Hu}, \citenamefont {Yang}, \citenamefont {Zhang}, \citenamefont {Li}, \citenamefont {Li}, \citenamefont {Jiang}, \citenamefont {Gan}, \citenamefont {Yang}, \citenamefont {You}, \citenamefont {Wang}, \citenamefont {Li}, \citenamefont {Liu}, \citenamefont {Lu},\ and\ \citenamefont {Pan}}]{Zhong2020Dec}%
  \BibitemOpen
  \bibfield  {author} {\bibinfo {author} {\bibfnamefont {H.-S.}\ \bibnamefont {Zhong}}, \bibinfo {author} {\bibfnamefont {H.}~\bibnamefont {Wang}}, \bibinfo {author} {\bibfnamefont {Y.-H.}\ \bibnamefont {Deng}}, \bibinfo {author} {\bibfnamefont {M.-C.}\ \bibnamefont {Chen}}, \bibinfo {author} {\bibfnamefont {L.-C.}\ \bibnamefont {Peng}}, \bibinfo {author} {\bibfnamefont {Y.-H.}\ \bibnamefont {Luo}}, \bibinfo {author} {\bibfnamefont {J.}~\bibnamefont {Qin}}, \bibinfo {author} {\bibfnamefont {D.}~\bibnamefont {Wu}}, \bibinfo {author} {\bibfnamefont {X.}~\bibnamefont {Ding}}, \bibinfo {author} {\bibfnamefont {Y.}~\bibnamefont {Hu}}, \bibinfo {author} {\bibfnamefont {P.}~\bibnamefont {Hu}}, \bibinfo {author} {\bibfnamefont {X.-Y.}\ \bibnamefont {Yang}}, \bibinfo {author} {\bibfnamefont {W.-J.}\ \bibnamefont {Zhang}}, \bibinfo {author} {\bibfnamefont {H.}~\bibnamefont {Li}}, \bibinfo {author} {\bibfnamefont {Y.}~\bibnamefont {Li}}, \bibinfo {author} {\bibfnamefont {X.}~\bibnamefont {Jiang}}, \bibinfo {author}
  {\bibfnamefont {L.}~\bibnamefont {Gan}}, \bibinfo {author} {\bibfnamefont {G.}~\bibnamefont {Yang}}, \bibinfo {author} {\bibfnamefont {L.}~\bibnamefont {You}}, \bibinfo {author} {\bibfnamefont {Z.}~\bibnamefont {Wang}}, \bibinfo {author} {\bibfnamefont {L.}~\bibnamefont {Li}}, \bibinfo {author} {\bibfnamefont {N.-L.}\ \bibnamefont {Liu}}, \bibinfo {author} {\bibfnamefont {C.-Y.}\ \bibnamefont {Lu}},\ and\ \bibinfo {author} {\bibfnamefont {J.-W.}\ \bibnamefont {Pan}},\ }\href {https://doi.org/10.1126/science.abe8770} {\bibfield  {journal} {\bibinfo  {journal} {Science}\ }\textbf {\bibinfo {volume} {370}},\ \bibinfo {pages} {1460} (\bibinfo {year} {2020})}\BibitemShut {NoStop}%
\bibitem [{\citenamefont {Senellart}\ \emph {et~al.}(2017)\citenamefont {Senellart}, \citenamefont {Solomon},\ and\ \citenamefont {White}}]{Senellart2017Nov}%
  \BibitemOpen
  \bibfield  {author} {\bibinfo {author} {\bibfnamefont {P.}~\bibnamefont {Senellart}}, \bibinfo {author} {\bibfnamefont {G.}~\bibnamefont {Solomon}},\ and\ \bibinfo {author} {\bibfnamefont {A.}~\bibnamefont {White}},\ }\href {https://doi.org/10.1038/nnano.2017.218} {\bibfield  {journal} {\bibinfo  {journal} {Nat. Nanotechnol.}\ }\textbf {\bibinfo {volume} {12}},\ \bibinfo {pages} {1026} (\bibinfo {year} {2017})}\BibitemShut {NoStop}%
\bibitem [{\citenamefont {Tomm}\ \emph {et~al.}(2021)\citenamefont {Tomm}, \citenamefont {Javadi}, \citenamefont {Antoniadis}, \citenamefont {Najer}, \citenamefont {L{\ifmmode\ddot{o}\else\"{o}\fi}bl}, \citenamefont {Korsch}, \citenamefont {Schott}, \citenamefont {Valentin}, \citenamefont {Wieck}, \citenamefont {Ludwig},\ and\ \citenamefont {Warburton}}]{Tomm2021Apr}%
  \BibitemOpen
  \bibfield  {author} {\bibinfo {author} {\bibfnamefont {N.}~\bibnamefont {Tomm}}, \bibinfo {author} {\bibfnamefont {A.}~\bibnamefont {Javadi}}, \bibinfo {author} {\bibfnamefont {N.~O.}\ \bibnamefont {Antoniadis}}, \bibinfo {author} {\bibfnamefont {D.}~\bibnamefont {Najer}}, \bibinfo {author} {\bibfnamefont {M.~C.}\ \bibnamefont {L{\ifmmode\ddot{o}\else\"{o}\fi}bl}}, \bibinfo {author} {\bibfnamefont {A.~R.}\ \bibnamefont {Korsch}}, \bibinfo {author} {\bibfnamefont {R.}~\bibnamefont {Schott}}, \bibinfo {author} {\bibfnamefont {S.~R.}\ \bibnamefont {Valentin}}, \bibinfo {author} {\bibfnamefont {A.~D.}\ \bibnamefont {Wieck}}, \bibinfo {author} {\bibfnamefont {A.}~\bibnamefont {Ludwig}},\ and\ \bibinfo {author} {\bibfnamefont {R.~J.}\ \bibnamefont {Warburton}},\ }\href {https://doi.org/10.1038/s41565-020-00831-x} {\bibfield  {journal} {\bibinfo  {journal} {Nat. Nanotechnol.}\ }\textbf {\bibinfo {volume} {16}},\ \bibinfo {pages} {399} (\bibinfo {year} {2021})}\BibitemShut {NoStop}%
\bibitem [{\citenamefont {Ding}\ \emph {et~al.}(2016)\citenamefont {Ding}, \citenamefont {He}, \citenamefont {Duan}, \citenamefont {Gregersen}, \citenamefont {Chen}, \citenamefont {Unsleber}, \citenamefont {Maier}, \citenamefont {Schneider}, \citenamefont {Kamp}, \citenamefont {H{\ifmmode\ddot{o}\else\"{o}\fi}fling}, \citenamefont {Lu},\ and\ \citenamefont {Pan}}]{Ding2016Jan}%
  \BibitemOpen
  \bibfield  {author} {\bibinfo {author} {\bibfnamefont {X.}~\bibnamefont {Ding}}, \bibinfo {author} {\bibfnamefont {Y.}~\bibnamefont {He}}, \bibinfo {author} {\bibfnamefont {Z.-C.}\ \bibnamefont {Duan}}, \bibinfo {author} {\bibfnamefont {N.}~\bibnamefont {Gregersen}}, \bibinfo {author} {\bibfnamefont {M.-C.}\ \bibnamefont {Chen}}, \bibinfo {author} {\bibfnamefont {S.}~\bibnamefont {Unsleber}}, \bibinfo {author} {\bibfnamefont {S.}~\bibnamefont {Maier}}, \bibinfo {author} {\bibfnamefont {C.}~\bibnamefont {Schneider}}, \bibinfo {author} {\bibfnamefont {M.}~\bibnamefont {Kamp}}, \bibinfo {author} {\bibfnamefont {S.}~\bibnamefont {H{\ifmmode\ddot{o}\else\"{o}\fi}fling}}, \bibinfo {author} {\bibfnamefont {C.-Y.}\ \bibnamefont {Lu}},\ and\ \bibinfo {author} {\bibfnamefont {J.-W.}\ \bibnamefont {Pan}},\ }\href {https://doi.org/10.1103/PhysRevLett.116.020401} {\bibfield  {journal} {\bibinfo  {journal} {Phys. Rev. Lett.}\ }\textbf {\bibinfo {volume} {116}},\ \bibinfo {pages} {020401} (\bibinfo {year}
  {2016})}\BibitemShut {NoStop}%
\bibitem [{\citenamefont {Mnaymneh}\ \emph {et~al.}(2020)\citenamefont {Mnaymneh}, \citenamefont {Dalacu}, \citenamefont {McKee}, \citenamefont {Lapointe}, \citenamefont {Haffouz}, \citenamefont {Weber}, \citenamefont {Northeast}, \citenamefont {Poole}, \citenamefont {Aers},\ and\ \citenamefont {Williams}}]{Mnaymneh2020Feb}%
  \BibitemOpen
  \bibfield  {author} {\bibinfo {author} {\bibfnamefont {K.}~\bibnamefont {Mnaymneh}}, \bibinfo {author} {\bibfnamefont {D.}~\bibnamefont {Dalacu}}, \bibinfo {author} {\bibfnamefont {J.}~\bibnamefont {McKee}}, \bibinfo {author} {\bibfnamefont {J.}~\bibnamefont {Lapointe}}, \bibinfo {author} {\bibfnamefont {S.}~\bibnamefont {Haffouz}}, \bibinfo {author} {\bibfnamefont {J.~F.}\ \bibnamefont {Weber}}, \bibinfo {author} {\bibfnamefont {D.~B.}\ \bibnamefont {Northeast}}, \bibinfo {author} {\bibfnamefont {P.~J.}\ \bibnamefont {Poole}}, \bibinfo {author} {\bibfnamefont {G.~C.}\ \bibnamefont {Aers}},\ and\ \bibinfo {author} {\bibfnamefont {R.~L.}\ \bibnamefont {Williams}},\ }\href {https://doi.org/10.1002/qute.201900021} {\bibfield  {journal} {\bibinfo  {journal} {Adv. Quantum Technol.}\ }\textbf {\bibinfo {volume} {3}},\ \bibinfo {pages} {1900021} (\bibinfo {year} {2020})}\BibitemShut {NoStop}%
\bibitem [{\citenamefont {Elshaari}\ \emph {et~al.}(2020)\citenamefont {Elshaari}, \citenamefont {Pernice}, \citenamefont {Srinivasan}, \citenamefont {Benson},\ and\ \citenamefont {Zwiller}}]{Elshaari2020May}%
  \BibitemOpen
  \bibfield  {author} {\bibinfo {author} {\bibfnamefont {A.~W.}\ \bibnamefont {Elshaari}}, \bibinfo {author} {\bibfnamefont {W.}~\bibnamefont {Pernice}}, \bibinfo {author} {\bibfnamefont {K.}~\bibnamefont {Srinivasan}}, \bibinfo {author} {\bibfnamefont {O.}~\bibnamefont {Benson}},\ and\ \bibinfo {author} {\bibfnamefont {V.}~\bibnamefont {Zwiller}},\ }\href {https://doi.org/10.1038/s41566-020-0609-x} {\bibfield  {journal} {\bibinfo  {journal} {Nat. Photonics}\ }\textbf {\bibinfo {volume} {14}},\ \bibinfo {pages} {285} (\bibinfo {year} {2020})}\BibitemShut {NoStop}%
\bibitem [{\citenamefont {Descamps}\ \emph {et~al.}(2023)\citenamefont {Descamps}, \citenamefont {Schetelat}, \citenamefont {Gao}, \citenamefont {Poole}, \citenamefont {Dalacu}, \citenamefont {Elshaari},\ and\ \citenamefont {Zwiller}}]{Descamps2023Oct}%
  \BibitemOpen
  \bibfield  {author} {\bibinfo {author} {\bibfnamefont {T.}~\bibnamefont {Descamps}}, \bibinfo {author} {\bibfnamefont {T.}~\bibnamefont {Schetelat}}, \bibinfo {author} {\bibfnamefont {J.}~\bibnamefont {Gao}}, \bibinfo {author} {\bibfnamefont {P.~J.}\ \bibnamefont {Poole}}, \bibinfo {author} {\bibfnamefont {D.}~\bibnamefont {Dalacu}}, \bibinfo {author} {\bibfnamefont {A.~W.}\ \bibnamefont {Elshaari}},\ and\ \bibinfo {author} {\bibfnamefont {V.}~\bibnamefont {Zwiller}},\ }\href {https://doi.org/10.1021/acsphotonics.3c00821} {\bibfield  {journal} {\bibinfo  {journal} {ACS Photonics}\ }\textbf {\bibinfo {volume} {10}},\ \bibinfo {pages} {3691} (\bibinfo {year} {2023})}\BibitemShut {NoStop}%
\bibitem [{\citenamefont {Yang}\ \emph {et~al.}(2024)\citenamefont {Yang}, \citenamefont {Jiang}, \citenamefont {Benthin}, \citenamefont {Hanel}, \citenamefont {Fandrich}, \citenamefont {Joos}, \citenamefont {Bauer}, \citenamefont {Kolatschek}, \citenamefont {Hreibi}, \citenamefont {Rugeramigabo}, \citenamefont {Jetter}, \citenamefont {Portalupi}, \citenamefont {Zopf}, \citenamefont {Michler}, \citenamefont {K{\ifmmode\ddot{u}\else\"{u}\fi}ck},\ and\ \citenamefont {Ding}}]{Yang2024Jul}%
  \BibitemOpen
  \bibfield  {author} {\bibinfo {author} {\bibfnamefont {J.}~\bibnamefont {Yang}}, \bibinfo {author} {\bibfnamefont {Z.}~\bibnamefont {Jiang}}, \bibinfo {author} {\bibfnamefont {F.}~\bibnamefont {Benthin}}, \bibinfo {author} {\bibfnamefont {J.}~\bibnamefont {Hanel}}, \bibinfo {author} {\bibfnamefont {T.}~\bibnamefont {Fandrich}}, \bibinfo {author} {\bibfnamefont {R.}~\bibnamefont {Joos}}, \bibinfo {author} {\bibfnamefont {S.}~\bibnamefont {Bauer}}, \bibinfo {author} {\bibfnamefont {S.}~\bibnamefont {Kolatschek}}, \bibinfo {author} {\bibfnamefont {A.}~\bibnamefont {Hreibi}}, \bibinfo {author} {\bibfnamefont {E.~P.}\ \bibnamefont {Rugeramigabo}}, \bibinfo {author} {\bibfnamefont {M.}~\bibnamefont {Jetter}}, \bibinfo {author} {\bibfnamefont {S.~L.}\ \bibnamefont {Portalupi}}, \bibinfo {author} {\bibfnamefont {M.}~\bibnamefont {Zopf}}, \bibinfo {author} {\bibfnamefont {P.}~\bibnamefont {Michler}}, \bibinfo {author} {\bibfnamefont {S.}~\bibnamefont {K{\ifmmode\ddot{u}\else\"{u}\fi}ck}},\ and\ \bibinfo {author}
  {\bibfnamefont {F.}~\bibnamefont {Ding}},\ }\href {https://doi.org/10.1038/s41377-024-01488-0} {\bibfield  {journal} {\bibinfo  {journal} {Light Sci. Appl.}\ }\textbf {\bibinfo {volume} {13}},\ \bibinfo {pages} {1} (\bibinfo {year} {2024})}\BibitemShut {NoStop}%
\bibitem [{\citenamefont {Zeuner}\ \emph {et~al.}(2021)\citenamefont {Zeuner}, \citenamefont {J{\ifmmode\ddot{o}\else\"{o}\fi}ns}, \citenamefont {Schweickert}, \citenamefont {Reuterski{\ifmmode\ddot{o}\else\"{o}\fi}ld~Hedlund}, \citenamefont {Nu{\ifmmode\tilde{n}\else\~{n}\fi}ez~Lobato}, \citenamefont {Lettner}, \citenamefont {Wang}, \citenamefont {Gyger}, \citenamefont {Sch{\ifmmode\ddot{o}\else\"{o}\fi}ll}, \citenamefont {Steinhauer}, \citenamefont {Hammar},\ and\ \citenamefont {Zwiller}}]{Zeuner2021Aug}%
  \BibitemOpen
  \bibfield  {author} {\bibinfo {author} {\bibfnamefont {K.~D.}\ \bibnamefont {Zeuner}}, \bibinfo {author} {\bibfnamefont {K.~D.}\ \bibnamefont {J{\ifmmode\ddot{o}\else\"{o}\fi}ns}}, \bibinfo {author} {\bibfnamefont {L.}~\bibnamefont {Schweickert}}, \bibinfo {author} {\bibfnamefont {C.}~\bibnamefont {Reuterski{\ifmmode\ddot{o}\else\"{o}\fi}ld~Hedlund}}, \bibinfo {author} {\bibfnamefont {C.}~\bibnamefont {Nu{\ifmmode\tilde{n}\else\~{n}\fi}ez~Lobato}}, \bibinfo {author} {\bibfnamefont {T.}~\bibnamefont {Lettner}}, \bibinfo {author} {\bibfnamefont {K.}~\bibnamefont {Wang}}, \bibinfo {author} {\bibfnamefont {S.}~\bibnamefont {Gyger}}, \bibinfo {author} {\bibfnamefont {E.}~\bibnamefont {Sch{\ifmmode\ddot{o}\else\"{o}\fi}ll}}, \bibinfo {author} {\bibfnamefont {S.}~\bibnamefont {Steinhauer}}, \bibinfo {author} {\bibfnamefont {M.}~\bibnamefont {Hammar}},\ and\ \bibinfo {author} {\bibfnamefont {V.}~\bibnamefont {Zwiller}},\ }\href {https://doi.org/10.1021/acsphotonics.1c00504} {\bibfield  {journal} {\bibinfo
  {journal} {ACS Photonics}\ }\textbf {\bibinfo {volume} {8}},\ \bibinfo {pages} {2337} (\bibinfo {year} {2021})}\BibitemShut {NoStop}%
\bibitem [{\citenamefont {Morrison}\ \emph {et~al.}(2023)\citenamefont {Morrison}, \citenamefont {Pousa}, \citenamefont {Graffitti}, \citenamefont {Koong}, \citenamefont {Barrow}, \citenamefont {Stoltz}, \citenamefont {Bouwmeester}, \citenamefont {Jeffers}, \citenamefont {Oi}, \citenamefont {Gerardot},\ and\ \citenamefont {Fedrizzi}}]{Morrison2023Jun}%
  \BibitemOpen
  \bibfield  {author} {\bibinfo {author} {\bibfnamefont {C.~L.}\ \bibnamefont {Morrison}}, \bibinfo {author} {\bibfnamefont {R.~G.}\ \bibnamefont {Pousa}}, \bibinfo {author} {\bibfnamefont {F.}~\bibnamefont {Graffitti}}, \bibinfo {author} {\bibfnamefont {Z.~X.}\ \bibnamefont {Koong}}, \bibinfo {author} {\bibfnamefont {P.}~\bibnamefont {Barrow}}, \bibinfo {author} {\bibfnamefont {N.~G.}\ \bibnamefont {Stoltz}}, \bibinfo {author} {\bibfnamefont {D.}~\bibnamefont {Bouwmeester}}, \bibinfo {author} {\bibfnamefont {J.}~\bibnamefont {Jeffers}}, \bibinfo {author} {\bibfnamefont {D.~K.~L.}\ \bibnamefont {Oi}}, \bibinfo {author} {\bibfnamefont {B.~D.}\ \bibnamefont {Gerardot}},\ and\ \bibinfo {author} {\bibfnamefont {A.}~\bibnamefont {Fedrizzi}},\ }\href {https://doi.org/10.1038/s41467-023-39219-5} {\bibfield  {journal} {\bibinfo  {journal} {Nat. Commun.}\ }\textbf {\bibinfo {volume} {14}},\ \bibinfo {pages} {1} (\bibinfo {year} {2023})}\BibitemShut {NoStop}%
\bibitem [{\citenamefont {Zahidy}\ \emph {et~al.}(2024)\citenamefont {Zahidy}, \citenamefont {Mikkelsen}, \citenamefont {M{\ifmmode\ddot{u}\else\"{u}\fi}ller}, \citenamefont {Da~Lio}, \citenamefont {Krehbiel}, \citenamefont {Wang}, \citenamefont {Bart}, \citenamefont {Wieck}, \citenamefont {Ludwig}, \citenamefont {Galili}, \citenamefont {Forchhammer}, \citenamefont {Lodahl}, \citenamefont {Oxenl{\o}we}, \citenamefont {Bacco},\ and\ \citenamefont {Midolo}}]{Zahidy2024Jan}%
  \BibitemOpen
  \bibfield  {author} {\bibinfo {author} {\bibfnamefont {M.}~\bibnamefont {Zahidy}}, \bibinfo {author} {\bibfnamefont {M.~T.}\ \bibnamefont {Mikkelsen}}, \bibinfo {author} {\bibfnamefont {R.}~\bibnamefont {M{\ifmmode\ddot{u}\else\"{u}\fi}ller}}, \bibinfo {author} {\bibfnamefont {B.}~\bibnamefont {Da~Lio}}, \bibinfo {author} {\bibfnamefont {M.}~\bibnamefont {Krehbiel}}, \bibinfo {author} {\bibfnamefont {Y.}~\bibnamefont {Wang}}, \bibinfo {author} {\bibfnamefont {N.}~\bibnamefont {Bart}}, \bibinfo {author} {\bibfnamefont {A.~D.}\ \bibnamefont {Wieck}}, \bibinfo {author} {\bibfnamefont {A.}~\bibnamefont {Ludwig}}, \bibinfo {author} {\bibfnamefont {M.}~\bibnamefont {Galili}}, \bibinfo {author} {\bibfnamefont {S.}~\bibnamefont {Forchhammer}}, \bibinfo {author} {\bibfnamefont {P.}~\bibnamefont {Lodahl}}, \bibinfo {author} {\bibfnamefont {L.~K.}\ \bibnamefont {Oxenl{\o}we}}, \bibinfo {author} {\bibfnamefont {D.}~\bibnamefont {Bacco}},\ and\ \bibinfo {author} {\bibfnamefont {L.}~\bibnamefont {Midolo}},\ }\href
  {https://doi.org/10.1038/s41534-023-00800-x} {\bibfield  {journal} {\bibinfo  {journal} {npj Quantum Inf.}\ }\textbf {\bibinfo {volume} {10}},\ \bibinfo {pages} {1} (\bibinfo {year} {2024})}\BibitemShut {NoStop}%
\bibitem [{\citenamefont {Miyazawa}\ \emph {et~al.}(2005)\citenamefont {Miyazawa}, \citenamefont {Takemoto}, \citenamefont {Sakuma}, \citenamefont {Hirose}, \citenamefont {Usuki}, \citenamefont {Yokoyama}, \citenamefont {Takatsu},\ and\ \citenamefont {Arakawa}}]{Miyazawa2005May}%
  \BibitemOpen
  \bibfield  {author} {\bibinfo {author} {\bibfnamefont {T.}~\bibnamefont {Miyazawa}}, \bibinfo {author} {\bibfnamefont {K.}~\bibnamefont {Takemoto}}, \bibinfo {author} {\bibfnamefont {Y.}~\bibnamefont {Sakuma}}, \bibinfo {author} {\bibfnamefont {S.}~\bibnamefont {Hirose}}, \bibinfo {author} {\bibfnamefont {T.}~\bibnamefont {Usuki}}, \bibinfo {author} {\bibfnamefont {N.}~\bibnamefont {Yokoyama}}, \bibinfo {author} {\bibfnamefont {M.}~\bibnamefont {Takatsu}},\ and\ \bibinfo {author} {\bibfnamefont {Y.}~\bibnamefont {Arakawa}},\ }\href {https://doi.org/10.1143/JJAP.44.L620} {\bibfield  {journal} {\bibinfo  {journal} {Jpn. J. Appl. Phys.}\ }\textbf {\bibinfo {volume} {44}},\ \bibinfo {pages} {L620} (\bibinfo {year} {2005})}\BibitemShut {NoStop}%
\bibitem [{\citenamefont {M{\ifmmode\ddot{u}\else\"{u}\fi}ller}\ \emph {et~al.}(2018)\citenamefont {M{\ifmmode\ddot{u}\else\"{u}\fi}ller}, \citenamefont {Skiba-Szymanska}, \citenamefont {Krysa}, \citenamefont {Huwer}, \citenamefont {Felle}, \citenamefont {Anderson}, \citenamefont {Stevenson}, \citenamefont {Heffernan}, \citenamefont {Ritchie},\ and\ \citenamefont {Shields}}]{Muller2018Feb}%
  \BibitemOpen
  \bibfield  {author} {\bibinfo {author} {\bibfnamefont {T.}~\bibnamefont {M{\ifmmode\ddot{u}\else\"{u}\fi}ller}}, \bibinfo {author} {\bibfnamefont {J.}~\bibnamefont {Skiba-Szymanska}}, \bibinfo {author} {\bibfnamefont {A.~B.}\ \bibnamefont {Krysa}}, \bibinfo {author} {\bibfnamefont {J.}~\bibnamefont {Huwer}}, \bibinfo {author} {\bibfnamefont {M.}~\bibnamefont {Felle}}, \bibinfo {author} {\bibfnamefont {M.}~\bibnamefont {Anderson}}, \bibinfo {author} {\bibfnamefont {R.~M.}\ \bibnamefont {Stevenson}}, \bibinfo {author} {\bibfnamefont {J.}~\bibnamefont {Heffernan}}, \bibinfo {author} {\bibfnamefont {D.~A.}\ \bibnamefont {Ritchie}},\ and\ \bibinfo {author} {\bibfnamefont {A.~J.}\ \bibnamefont {Shields}},\ }\href {https://doi.org/10.1038/s41467-018-03251-7} {\bibfield  {journal} {\bibinfo  {journal} {Nat. Commun.}\ }\textbf {\bibinfo {volume} {9}},\ \bibinfo {pages} {1} (\bibinfo {year} {2018})}\BibitemShut {NoStop}%
\bibitem [{\citenamefont {Nawrath}\ \emph {et~al.}(2019)\citenamefont {Nawrath}, \citenamefont {Olbrich}, \citenamefont {Paul}, \citenamefont {Portalupi}, \citenamefont {Jetter},\ and\ \citenamefont {Michler}}]{Nawrath2019Jul}%
  \BibitemOpen
  \bibfield  {author} {\bibinfo {author} {\bibfnamefont {C.}~\bibnamefont {Nawrath}}, \bibinfo {author} {\bibfnamefont {F.}~\bibnamefont {Olbrich}}, \bibinfo {author} {\bibfnamefont {M.}~\bibnamefont {Paul}}, \bibinfo {author} {\bibfnamefont {S.~L.}\ \bibnamefont {Portalupi}}, \bibinfo {author} {\bibfnamefont {M.}~\bibnamefont {Jetter}},\ and\ \bibinfo {author} {\bibfnamefont {P.}~\bibnamefont {Michler}},\ }\bibfield  {journal} {\bibinfo  {journal} {Appl. Phys. Lett.}\ }\textbf {\bibinfo {volume} {115}},\ \href {https://doi.org/10.1063/1.5095196} {10.1063/1.5095196} (\bibinfo {year} {2019})\BibitemShut {NoStop}%
\bibitem [{\citenamefont {Vajner}\ \emph {et~al.}(2024)\citenamefont {Vajner}, \citenamefont {Holewa}, \citenamefont {Zi{\k{e}}ba-Ost{\ifmmode\acute{o}\else\'{o}\fi}j}, \citenamefont {Wasiluk}, \citenamefont {von Helversen}, \citenamefont {Sakanas}, \citenamefont {Huck}, \citenamefont {Yvind}, \citenamefont {Gregersen}, \citenamefont {Musia{\l}}, \citenamefont {Syperek}, \citenamefont {Semenova},\ and\ \citenamefont {Heindel}}]{Vajner2024Feb}%
  \BibitemOpen
  \bibfield  {author} {\bibinfo {author} {\bibfnamefont {D.~A.}\ \bibnamefont {Vajner}}, \bibinfo {author} {\bibfnamefont {P.}~\bibnamefont {Holewa}}, \bibinfo {author} {\bibfnamefont {E.}~\bibnamefont {Zi{\k{e}}ba-Ost{\ifmmode\acute{o}\else\'{o}\fi}j}}, \bibinfo {author} {\bibfnamefont {M.}~\bibnamefont {Wasiluk}}, \bibinfo {author} {\bibfnamefont {M.}~\bibnamefont {von Helversen}}, \bibinfo {author} {\bibfnamefont {A.}~\bibnamefont {Sakanas}}, \bibinfo {author} {\bibfnamefont {A.}~\bibnamefont {Huck}}, \bibinfo {author} {\bibfnamefont {K.}~\bibnamefont {Yvind}}, \bibinfo {author} {\bibfnamefont {N.}~\bibnamefont {Gregersen}}, \bibinfo {author} {\bibfnamefont {A.}~\bibnamefont {Musia{\l}}}, \bibinfo {author} {\bibfnamefont {M.}~\bibnamefont {Syperek}}, \bibinfo {author} {\bibfnamefont {E.}~\bibnamefont {Semenova}},\ and\ \bibinfo {author} {\bibfnamefont {T.}~\bibnamefont {Heindel}},\ }\href {https://doi.org/10.1021/acsphotonics.3c00973} {\bibfield  {journal} {\bibinfo  {journal} {ACS Photonics}\ }\textbf
  {\bibinfo {volume} {11}},\ \bibinfo {pages} {339} (\bibinfo {year} {2024})}\BibitemShut {NoStop}%
\bibitem [{\citenamefont {Rahaman}\ \emph {et~al.}(2024)\citenamefont {Rahaman}, \citenamefont {Harper}, \citenamefont {Lee}, \citenamefont {Kim}, \citenamefont {Buyukkaya}, \citenamefont {Patel}, \citenamefont {Hawkins}, \citenamefont {Kim}, \citenamefont {Addamane},\ and\ \citenamefont {Waks}}]{Rahaman2024Jul}%
  \BibitemOpen
  \bibfield  {author} {\bibinfo {author} {\bibfnamefont {M.~H.}\ \bibnamefont {Rahaman}}, \bibinfo {author} {\bibfnamefont {S.}~\bibnamefont {Harper}}, \bibinfo {author} {\bibfnamefont {C.-M.}\ \bibnamefont {Lee}}, \bibinfo {author} {\bibfnamefont {K.-Y.}\ \bibnamefont {Kim}}, \bibinfo {author} {\bibfnamefont {M.~A.}\ \bibnamefont {Buyukkaya}}, \bibinfo {author} {\bibfnamefont {V.~J.}\ \bibnamefont {Patel}}, \bibinfo {author} {\bibfnamefont {S.~D.}\ \bibnamefont {Hawkins}}, \bibinfo {author} {\bibfnamefont {J.-H.}\ \bibnamefont {Kim}}, \bibinfo {author} {\bibfnamefont {S.}~\bibnamefont {Addamane}},\ and\ \bibinfo {author} {\bibfnamefont {E.}~\bibnamefont {Waks}},\ }\href {https://doi.org/10.1021/acsphotonics.4c00625} {\bibfield  {journal} {\bibinfo  {journal} {ACS Photonics}\ }\textbf {\bibinfo {volume} {11}},\ \bibinfo {pages} {2738} (\bibinfo {year} {2024})}\BibitemShut {NoStop}%
\bibitem [{\citenamefont {Joos}\ \emph {et~al.}(2024)\citenamefont {Joos}, \citenamefont {Bauer}, \citenamefont {Rupp}, \citenamefont {Kolatschek}, \citenamefont {Fischer}, \citenamefont {Nawrath}, \citenamefont {Vijayan}, \citenamefont {Sittig}, \citenamefont {Jetter}, \citenamefont {Portalupi},\ and\ \citenamefont {Michler}}]{Joos2024Jul}%
  \BibitemOpen
  \bibfield  {author} {\bibinfo {author} {\bibfnamefont {R.}~\bibnamefont {Joos}}, \bibinfo {author} {\bibfnamefont {S.}~\bibnamefont {Bauer}}, \bibinfo {author} {\bibfnamefont {C.}~\bibnamefont {Rupp}}, \bibinfo {author} {\bibfnamefont {S.}~\bibnamefont {Kolatschek}}, \bibinfo {author} {\bibfnamefont {W.}~\bibnamefont {Fischer}}, \bibinfo {author} {\bibfnamefont {C.}~\bibnamefont {Nawrath}}, \bibinfo {author} {\bibfnamefont {P.}~\bibnamefont {Vijayan}}, \bibinfo {author} {\bibfnamefont {R.}~\bibnamefont {Sittig}}, \bibinfo {author} {\bibfnamefont {M.}~\bibnamefont {Jetter}}, \bibinfo {author} {\bibfnamefont {S.~L.}\ \bibnamefont {Portalupi}},\ and\ \bibinfo {author} {\bibfnamefont {P.}~\bibnamefont {Michler}},\ }\href {https://doi.org/10.1021/acs.nanolett.4c01813} {\bibfield  {journal} {\bibinfo  {journal} {Nano Lett.}\ }\textbf {\bibinfo {volume} {24}},\ \bibinfo {pages} {8626} (\bibinfo {year} {2024})}\BibitemShut {NoStop}%
\bibitem [{\citenamefont {Ge}\ \emph {et~al.}(2024)\citenamefont {Ge}, \citenamefont {Chung}, \citenamefont {He}, \citenamefont {Benyoucef},\ and\ \citenamefont {Huo}}]{Ge2024Feb}%
  \BibitemOpen
  \bibfield  {author} {\bibinfo {author} {\bibfnamefont {Z.}~\bibnamefont {Ge}}, \bibinfo {author} {\bibfnamefont {T.}~\bibnamefont {Chung}}, \bibinfo {author} {\bibfnamefont {Y.-M.}\ \bibnamefont {He}}, \bibinfo {author} {\bibfnamefont {M.}~\bibnamefont {Benyoucef}},\ and\ \bibinfo {author} {\bibfnamefont {Y.}~\bibnamefont {Huo}},\ }\href {https://doi.org/10.1021/acs.nanolett.3c04618} {\bibfield  {journal} {\bibinfo  {journal} {Nano Lett.}\ }\textbf {\bibinfo {volume} {24}},\ \bibinfo {pages} {1746} (\bibinfo {year} {2024})}\BibitemShut {NoStop}%
\bibitem [{\citenamefont {Holewa}\ \emph {et~al.}(2024)\citenamefont {Holewa}, \citenamefont {Vajner}, \citenamefont {Zi{\k{e}}ba-Ost{\ifmmode\acute{o}\else\'{o}\fi}j}, \citenamefont {Wasiluk}, \citenamefont {Ga{\ifmmode\acute{a}\else\'{a}\fi}l}, \citenamefont {Sakanas}, \citenamefont {Burakowski}, \citenamefont {Mrowi{\ifmmode\acute{n}\else\'{n}\fi}ski}, \citenamefont {Krajnik}, \citenamefont {Xiong}, \citenamefont {Yvind}, \citenamefont {Gregersen}, \citenamefont {Musia{\l}}, \citenamefont {Huck}, \citenamefont {Heindel}, \citenamefont {Syperek},\ and\ \citenamefont {Semenova}}]{Holewa2024Apr}%
  \BibitemOpen
  \bibfield  {author} {\bibinfo {author} {\bibfnamefont {P.}~\bibnamefont {Holewa}}, \bibinfo {author} {\bibfnamefont {D.~A.}\ \bibnamefont {Vajner}}, \bibinfo {author} {\bibfnamefont {E.}~\bibnamefont {Zi{\k{e}}ba-Ost{\ifmmode\acute{o}\else\'{o}\fi}j}}, \bibinfo {author} {\bibfnamefont {M.}~\bibnamefont {Wasiluk}}, \bibinfo {author} {\bibfnamefont {B.}~\bibnamefont {Ga{\ifmmode\acute{a}\else\'{a}\fi}l}}, \bibinfo {author} {\bibfnamefont {A.}~\bibnamefont {Sakanas}}, \bibinfo {author} {\bibfnamefont {M.}~\bibnamefont {Burakowski}}, \bibinfo {author} {\bibfnamefont {P.}~\bibnamefont {Mrowi{\ifmmode\acute{n}\else\'{n}\fi}ski}}, \bibinfo {author} {\bibfnamefont {B.}~\bibnamefont {Krajnik}}, \bibinfo {author} {\bibfnamefont {M.}~\bibnamefont {Xiong}}, \bibinfo {author} {\bibfnamefont {K.}~\bibnamefont {Yvind}}, \bibinfo {author} {\bibfnamefont {N.}~\bibnamefont {Gregersen}}, \bibinfo {author} {\bibfnamefont {A.}~\bibnamefont {Musia{\l}}}, \bibinfo {author} {\bibfnamefont {A.}~\bibnamefont {Huck}}, \bibinfo
  {author} {\bibfnamefont {T.}~\bibnamefont {Heindel}}, \bibinfo {author} {\bibfnamefont {M.}~\bibnamefont {Syperek}},\ and\ \bibinfo {author} {\bibfnamefont {E.}~\bibnamefont {Semenova}},\ }\href {https://doi.org/10.1038/s41467-024-47551-7} {\bibfield  {journal} {\bibinfo  {journal} {Nat. Commun.}\ }\textbf {\bibinfo {volume} {15}},\ \bibinfo {pages} {1} (\bibinfo {year} {2024})}\BibitemShut {NoStop}%
\bibitem [{\citenamefont {Laferri{\ifmmode\grave{e}\else\`{e}\fi}re}\ \emph {et~al.}(2022)\citenamefont {Laferri{\ifmmode\grave{e}\else\`{e}\fi}re}, \citenamefont {Yeung}, \citenamefont {Miron}, \citenamefont {Northeast}, \citenamefont {Haffouz}, \citenamefont {Lapointe}, \citenamefont {Korkusinski}, \citenamefont {Poole}, \citenamefont {Williams},\ and\ \citenamefont {Dalacu}}]{Laferriere2022Apr}%
  \BibitemOpen
  \bibfield  {author} {\bibinfo {author} {\bibfnamefont {P.}~\bibnamefont {Laferri{\ifmmode\grave{e}\else\`{e}\fi}re}}, \bibinfo {author} {\bibfnamefont {E.}~\bibnamefont {Yeung}}, \bibinfo {author} {\bibfnamefont {I.}~\bibnamefont {Miron}}, \bibinfo {author} {\bibfnamefont {D.~B.}\ \bibnamefont {Northeast}}, \bibinfo {author} {\bibfnamefont {S.}~\bibnamefont {Haffouz}}, \bibinfo {author} {\bibfnamefont {J.}~\bibnamefont {Lapointe}}, \bibinfo {author} {\bibfnamefont {M.}~\bibnamefont {Korkusinski}}, \bibinfo {author} {\bibfnamefont {P.~J.}\ \bibnamefont {Poole}}, \bibinfo {author} {\bibfnamefont {R.~L.}\ \bibnamefont {Williams}},\ and\ \bibinfo {author} {\bibfnamefont {D.}~\bibnamefont {Dalacu}},\ }\href {https://doi.org/10.1038/s41598-022-10451-1} {\bibfield  {journal} {\bibinfo  {journal} {Sci. Rep.}\ }\textbf {\bibinfo {volume} {12}},\ \bibinfo {pages} {1} (\bibinfo {year} {2022})}\BibitemShut {NoStop}%
\bibitem [{\citenamefont {Wakileh}\ \emph {et~al.}(2024)\citenamefont {Wakileh}, \citenamefont {Yu}, \citenamefont {Dokuz}, \citenamefont {Haffouz}, \citenamefont {Wu}, \citenamefont {Lapointe}, \citenamefont {Northeast}, \citenamefont {Williams}, \citenamefont {Rotenberg}, \citenamefont {Poole},\ and\ \citenamefont {Dalacu}}]{Wakileh2024Jan}%
  \BibitemOpen
  \bibfield  {author} {\bibinfo {author} {\bibfnamefont {A.~N.}\ \bibnamefont {Wakileh}}, \bibinfo {author} {\bibfnamefont {L.}~\bibnamefont {Yu}}, \bibinfo {author} {\bibfnamefont {D.}~\bibnamefont {Dokuz}}, \bibinfo {author} {\bibfnamefont {S.}~\bibnamefont {Haffouz}}, \bibinfo {author} {\bibfnamefont {X.}~\bibnamefont {Wu}}, \bibinfo {author} {\bibfnamefont {J.}~\bibnamefont {Lapointe}}, \bibinfo {author} {\bibfnamefont {D.~B.}\ \bibnamefont {Northeast}}, \bibinfo {author} {\bibfnamefont {R.~L.}\ \bibnamefont {Williams}}, \bibinfo {author} {\bibfnamefont {N.}~\bibnamefont {Rotenberg}}, \bibinfo {author} {\bibfnamefont {P.~J.}\ \bibnamefont {Poole}},\ and\ \bibinfo {author} {\bibfnamefont {D.}~\bibnamefont {Dalacu}},\ }\href {https://doi.org/10.1063/5.0179234} {\bibfield  {journal} {\bibinfo  {journal} {Appl. Phys. Lett.}\ }\textbf {\bibinfo {volume} {124}},\ \bibinfo {pages} {044006} (\bibinfo {year} {2024})}\BibitemShut {NoStop}%
\bibitem [{\citenamefont {Alqedra}\ \emph {et~al.}(2025)\citenamefont {Alqedra}, \citenamefont {Huang}, \citenamefont {Yeung}, \citenamefont {Chang}, \citenamefont {Haffouz}, \citenamefont {Poole}, \citenamefont {Dalacu}, \citenamefont {Elshaari},\ and\ \citenamefont {Zwiller}}]{Alqedra2025Feb}%
  \BibitemOpen
  \bibfield  {author} {\bibinfo {author} {\bibfnamefont {M.~K.}\ \bibnamefont {Alqedra}}, \bibinfo {author} {\bibfnamefont {C.-T.}\ \bibnamefont {Huang}}, \bibinfo {author} {\bibfnamefont {E.}~\bibnamefont {Yeung}}, \bibinfo {author} {\bibfnamefont {W.-H.}\ \bibnamefont {Chang}}, \bibinfo {author} {\bibfnamefont {S.}~\bibnamefont {Haffouz}}, \bibinfo {author} {\bibfnamefont {P.~J.}\ \bibnamefont {Poole}}, \bibinfo {author} {\bibfnamefont {D.}~\bibnamefont {Dalacu}}, \bibinfo {author} {\bibfnamefont {A.~W.}\ \bibnamefont {Elshaari}},\ and\ \bibinfo {author} {\bibfnamefont {V.}~\bibnamefont {Zwiller}},\ }\bibfield  {journal} {\bibinfo  {journal} {arXiv}\ }\href {https://doi.org/10.48550/arXiv.2502.14071} {10.48550/arXiv.2502.14071} (\bibinfo {year} {2025}),\ \Eprint {https://arxiv.org/abs/2502.14071} {2502.14071} \BibitemShut {NoStop}%
\bibitem [{\citenamefont {Chang}\ \emph {et~al.}(2023)\citenamefont {Chang}, \citenamefont {Gao}, \citenamefont {Zadeh}, \citenamefont {Elshaari},\ and\ \citenamefont {Zwiller}}]{Chang2023Feb}%
  \BibitemOpen
  \bibfield  {author} {\bibinfo {author} {\bibfnamefont {J.}~\bibnamefont {Chang}}, \bibinfo {author} {\bibfnamefont {J.}~\bibnamefont {Gao}}, \bibinfo {author} {\bibfnamefont {I.~E.}\ \bibnamefont {Zadeh}}, \bibinfo {author} {\bibfnamefont {A.~W.}\ \bibnamefont {Elshaari}},\ and\ \bibinfo {author} {\bibfnamefont {V.}~\bibnamefont {Zwiller}},\ }\href {https://doi.org/10.1515/nanoph-2022-0652} {\bibfield  {journal} {\bibinfo  {journal} {Nanophotonics}\ }\textbf {\bibinfo {volume} {12}},\ \bibinfo {pages} {339} (\bibinfo {year} {2023})}\BibitemShut {NoStop}%
\bibitem [{\citenamefont {Descamps}\ \emph {et~al.}(2024)\citenamefont {Descamps}, \citenamefont {Schetelat}, \citenamefont {Gao}, \citenamefont {Poole}, \citenamefont {Dalacu}, \citenamefont {Elshaari},\ and\ \citenamefont {Zwiller}}]{Descamps2024Oct}%
  \BibitemOpen
  \bibfield  {author} {\bibinfo {author} {\bibfnamefont {T.}~\bibnamefont {Descamps}}, \bibinfo {author} {\bibfnamefont {T.}~\bibnamefont {Schetelat}}, \bibinfo {author} {\bibfnamefont {J.}~\bibnamefont {Gao}}, \bibinfo {author} {\bibfnamefont {P.~J.}\ \bibnamefont {Poole}}, \bibinfo {author} {\bibfnamefont {D.}~\bibnamefont {Dalacu}}, \bibinfo {author} {\bibfnamefont {A.~W.}\ \bibnamefont {Elshaari}},\ and\ \bibinfo {author} {\bibfnamefont {V.}~\bibnamefont {Zwiller}},\ }\href {https://doi.org/10.1021/acs.nanolett.4c03402} {\bibfield  {journal} {\bibinfo  {journal} {Nano Lett.}\ }\textbf {\bibinfo {volume} {24}},\ \bibinfo {pages} {12493} (\bibinfo {year} {2024})}\BibitemShut {NoStop}%
\bibitem [{\citenamefont {Dalacu}\ \emph {et~al.}(2011)\citenamefont {Dalacu}, \citenamefont {Mnaymneh}, \citenamefont {Wu}, \citenamefont {Lapointe}, \citenamefont {Aers}, \citenamefont {Poole},\ and\ \citenamefont {Williams}}]{Dalacu2011Jun}%
  \BibitemOpen
  \bibfield  {author} {\bibinfo {author} {\bibfnamefont {D.}~\bibnamefont {Dalacu}}, \bibinfo {author} {\bibfnamefont {K.}~\bibnamefont {Mnaymneh}}, \bibinfo {author} {\bibfnamefont {X.}~\bibnamefont {Wu}}, \bibinfo {author} {\bibfnamefont {J.}~\bibnamefont {Lapointe}}, \bibinfo {author} {\bibfnamefont {G.~C.}\ \bibnamefont {Aers}}, \bibinfo {author} {\bibfnamefont {P.~J.}\ \bibnamefont {Poole}},\ and\ \bibinfo {author} {\bibfnamefont {R.~L.}\ \bibnamefont {Williams}},\ }\href {https://doi.org/10.1063/1.3600777} {\bibfield  {journal} {\bibinfo  {journal} {Appl. Phys. Lett.}\ }\textbf {\bibinfo {volume} {98}},\ \bibinfo {pages} {251101} (\bibinfo {year} {2011})}\BibitemShut {NoStop}%
\bibitem [{\citenamefont {Haffouz}\ \emph {et~al.}(2020)\citenamefont {Haffouz}, \citenamefont {Poole}, \citenamefont {Jin}, \citenamefont {Wu}, \citenamefont {Ginet}, \citenamefont {Mnaymneh}, \citenamefont {Dalacu},\ and\ \citenamefont {Williams}}]{Haffouz2020Sep}%
  \BibitemOpen
  \bibfield  {author} {\bibinfo {author} {\bibfnamefont {S.}~\bibnamefont {Haffouz}}, \bibinfo {author} {\bibfnamefont {P.~J.}\ \bibnamefont {Poole}}, \bibinfo {author} {\bibfnamefont {J.}~\bibnamefont {Jin}}, \bibinfo {author} {\bibfnamefont {X.}~\bibnamefont {Wu}}, \bibinfo {author} {\bibfnamefont {L.}~\bibnamefont {Ginet}}, \bibinfo {author} {\bibfnamefont {K.}~\bibnamefont {Mnaymneh}}, \bibinfo {author} {\bibfnamefont {D.}~\bibnamefont {Dalacu}},\ and\ \bibinfo {author} {\bibfnamefont {R.~L.}\ \bibnamefont {Williams}},\ }\href {https://doi.org/10.1063/5.0020681} {\bibfield  {journal} {\bibinfo  {journal} {Appl. Phys. Lett.}\ }\textbf {\bibinfo {volume} {117}},\ \bibinfo {pages} {113102} (\bibinfo {year} {2020})}\BibitemShut {NoStop}%
\bibitem [{\citenamefont {Laferri{\ifmmode\acute{e}\else\'{e}\fi}re}\ \emph {et~al.}(2023)\citenamefont {Laferri{\ifmmode\acute{e}\else\'{e}\fi}re}, \citenamefont {Haffouz}, \citenamefont {Northeast}, \citenamefont {Poole}, \citenamefont {Williams},\ and\ \citenamefont {Dalacu}}]{Laferriere2023Feb}%
  \BibitemOpen
  \bibfield  {author} {\bibinfo {author} {\bibfnamefont {P.}~\bibnamefont {Laferri{\ifmmode\acute{e}\else\'{e}\fi}re}}, \bibinfo {author} {\bibfnamefont {S.}~\bibnamefont {Haffouz}}, \bibinfo {author} {\bibfnamefont {D.~B.}\ \bibnamefont {Northeast}}, \bibinfo {author} {\bibfnamefont {P.~J.}\ \bibnamefont {Poole}}, \bibinfo {author} {\bibfnamefont {R.~L.}\ \bibnamefont {Williams}},\ and\ \bibinfo {author} {\bibfnamefont {D.}~\bibnamefont {Dalacu}},\ }\href {https://doi.org/10.1021/acs.nanolett.2c04375} {\bibfield  {journal} {\bibinfo  {journal} {Nano Lett.}\ }\textbf {\bibinfo {volume} {23}},\ \bibinfo {pages} {962} (\bibinfo {year} {2023})}\BibitemShut {NoStop}%
\bibitem [{\citenamefont {Lin}\ \emph {et~al.}(2021)\citenamefont {Lin}, \citenamefont {Schweickert}, \citenamefont {Gyger}, \citenamefont {J{\ifmmode\ddot{o}\else\"{o}\fi}ns},\ and\ \citenamefont {Zwiller}}]{Lin2021Aug}%
  \BibitemOpen
  \bibfield  {author} {\bibinfo {author} {\bibfnamefont {Z.}~\bibnamefont {Lin}}, \bibinfo {author} {\bibfnamefont {L.}~\bibnamefont {Schweickert}}, \bibinfo {author} {\bibfnamefont {S.}~\bibnamefont {Gyger}}, \bibinfo {author} {\bibfnamefont {K.~D.}\ \bibnamefont {J{\ifmmode\ddot{o}\else\"{o}\fi}ns}},\ and\ \bibinfo {author} {\bibfnamefont {V.}~\bibnamefont {Zwiller}},\ }\href {https://doi.org/10.1088/1748-0221/16/08/T08016} {\bibfield  {journal} {\bibinfo  {journal} {J. Instrum.}\ }\textbf {\bibinfo {volume} {16}}\bibinfo  {number} { (08)},\ \bibinfo {pages} {T08016}}\BibitemShut {NoStop}%
\bibitem [{\citenamefont {Laferri{\ifmmode\grave{e}\else\`{e}\fi}re}\ \emph {et~al.}(2023)\citenamefont {Laferri{\ifmmode\grave{e}\else\`{e}\fi}re}, \citenamefont {Yin}, \citenamefont {Yeung}, \citenamefont {Kusmic}, \citenamefont {Korkusinski}, \citenamefont {Rasekh}, \citenamefont {Northeast}, \citenamefont {Haffouz}, \citenamefont {Lapointe}, \citenamefont {Poole}, \citenamefont {Williams},\ and\ \citenamefont {Dalacu}}]{Laferriere2023Apr}%
  \BibitemOpen
\bibfield  {number} {  }\bibfield  {author} {\bibinfo {author} {\bibfnamefont {P.}~\bibnamefont {Laferri{\ifmmode\grave{e}\else\`{e}\fi}re}}, \bibinfo {author} {\bibfnamefont {A.}~\bibnamefont {Yin}}, \bibinfo {author} {\bibfnamefont {E.}~\bibnamefont {Yeung}}, \bibinfo {author} {\bibfnamefont {L.}~\bibnamefont {Kusmic}}, \bibinfo {author} {\bibfnamefont {M.}~\bibnamefont {Korkusinski}}, \bibinfo {author} {\bibfnamefont {P.}~\bibnamefont {Rasekh}}, \bibinfo {author} {\bibfnamefont {D.~B.}\ \bibnamefont {Northeast}}, \bibinfo {author} {\bibfnamefont {S.}~\bibnamefont {Haffouz}}, \bibinfo {author} {\bibfnamefont {J.}~\bibnamefont {Lapointe}}, \bibinfo {author} {\bibfnamefont {P.~J.}\ \bibnamefont {Poole}}, \bibinfo {author} {\bibfnamefont {R.~L.}\ \bibnamefont {Williams}},\ and\ \bibinfo {author} {\bibfnamefont {D.}~\bibnamefont {Dalacu}},\ }\href {https://doi.org/10.1103/PhysRevB.107.155422} {\bibfield  {journal} {\bibinfo  {journal} {Phys. Rev. B}\ }\textbf {\bibinfo {volume} {107}},\ \bibinfo {pages}
  {155422} (\bibinfo {year} {2023})}\BibitemShut {NoStop}%
\bibitem [{\citenamefont {L{\ifmmode\ddot{o}\else\"{o}\fi}bl}\ \emph {et~al.}(2019)\citenamefont {L{\ifmmode\ddot{o}\else\"{o}\fi}bl}, \citenamefont {Scholz}, \citenamefont {S{\ifmmode\ddot{o}\else\"{o}\fi}llner}, \citenamefont {Ritzmann}, \citenamefont {Denneulin}, \citenamefont {Kov{\ifmmode\acute{a}\else\'{a}\fi}cs}, \citenamefont {Kardyna{\l}}, \citenamefont {Wieck}, \citenamefont {Ludwig},\ and\ \citenamefont {Warburton}}]{Lobl2019Aug}%
  \BibitemOpen
  \bibfield  {author} {\bibinfo {author} {\bibfnamefont {M.~C.}\ \bibnamefont {L{\ifmmode\ddot{o}\else\"{o}\fi}bl}}, \bibinfo {author} {\bibfnamefont {S.}~\bibnamefont {Scholz}}, \bibinfo {author} {\bibfnamefont {I.}~\bibnamefont {S{\ifmmode\ddot{o}\else\"{o}\fi}llner}}, \bibinfo {author} {\bibfnamefont {J.}~\bibnamefont {Ritzmann}}, \bibinfo {author} {\bibfnamefont {T.}~\bibnamefont {Denneulin}}, \bibinfo {author} {\bibfnamefont {A.}~\bibnamefont {Kov{\ifmmode\acute{a}\else\'{a}\fi}cs}}, \bibinfo {author} {\bibfnamefont {B.~E.}\ \bibnamefont {Kardyna{\l}}}, \bibinfo {author} {\bibfnamefont {A.~D.}\ \bibnamefont {Wieck}}, \bibinfo {author} {\bibfnamefont {A.}~\bibnamefont {Ludwig}},\ and\ \bibinfo {author} {\bibfnamefont {R.~J.}\ \bibnamefont {Warburton}},\ }\href {https://doi.org/10.1038/s42005-019-0194-9} {\bibfield  {journal} {\bibinfo  {journal} {Commun. Phys.}\ }\textbf {\bibinfo {volume} {2}},\ \bibinfo {pages} {1} (\bibinfo {year} {2019})}\BibitemShut {NoStop}%
\bibitem [{\citenamefont {Urbaszek}\ \emph {et~al.}(2004)\citenamefont {Urbaszek}, \citenamefont {McGhee}, \citenamefont {Kr{\ifmmode\ddot{u}\else\"{u}\fi}ger}, \citenamefont {Warburton}, \citenamefont {Karrai}, \citenamefont {Amand}, \citenamefont {Gerardot}, \citenamefont {Petroff},\ and\ \citenamefont {Garcia}}]{Urbaszek2004Jan}%
  \BibitemOpen
  \bibfield  {author} {\bibinfo {author} {\bibfnamefont {B.}~\bibnamefont {Urbaszek}}, \bibinfo {author} {\bibfnamefont {E.~J.}\ \bibnamefont {McGhee}}, \bibinfo {author} {\bibfnamefont {M.}~\bibnamefont {Kr{\ifmmode\ddot{u}\else\"{u}\fi}ger}}, \bibinfo {author} {\bibfnamefont {R.~J.}\ \bibnamefont {Warburton}}, \bibinfo {author} {\bibfnamefont {K.}~\bibnamefont {Karrai}}, \bibinfo {author} {\bibfnamefont {T.}~\bibnamefont {Amand}}, \bibinfo {author} {\bibfnamefont {B.~D.}\ \bibnamefont {Gerardot}}, \bibinfo {author} {\bibfnamefont {P.~M.}\ \bibnamefont {Petroff}},\ and\ \bibinfo {author} {\bibfnamefont {J.~M.}\ \bibnamefont {Garcia}},\ }\href {https://doi.org/10.1103/PhysRevB.69.035304} {\bibfield  {journal} {\bibinfo  {journal} {Phys. Rev. B}\ }\textbf {\bibinfo {volume} {69}},\ \bibinfo {pages} {035304} (\bibinfo {year} {2004})}\BibitemShut {NoStop}%
\bibitem [{\citenamefont {Gao}\ \emph {et~al.}(2024{\natexlab{a}})\citenamefont {Gao}, \citenamefont {Krishna}, \citenamefont {Yeung}, \citenamefont {Yu}, \citenamefont {Gangopadhyay}, \citenamefont {Chan}, \citenamefont {Huang}, \citenamefont {Descamps}, \citenamefont {Reimer}, \citenamefont {Poole}, \citenamefont {Dalacu}, \citenamefont {Zwiller},\ and\ \citenamefont {Elshaari}}]{Gao2024Sep}%
  \BibitemOpen
  \bibfield  {author} {\bibinfo {author} {\bibfnamefont {J.}~\bibnamefont {Gao}}, \bibinfo {author} {\bibfnamefont {G.}~\bibnamefont {Krishna}}, \bibinfo {author} {\bibfnamefont {E.}~\bibnamefont {Yeung}}, \bibinfo {author} {\bibfnamefont {L.}~\bibnamefont {Yu}}, \bibinfo {author} {\bibfnamefont {S.}~\bibnamefont {Gangopadhyay}}, \bibinfo {author} {\bibfnamefont {K.-S.}\ \bibnamefont {Chan}}, \bibinfo {author} {\bibfnamefont {C.-T.}\ \bibnamefont {Huang}}, \bibinfo {author} {\bibfnamefont {T.}~\bibnamefont {Descamps}}, \bibinfo {author} {\bibfnamefont {M.~E.}\ \bibnamefont {Reimer}}, \bibinfo {author} {\bibfnamefont {P.~J.}\ \bibnamefont {Poole}}, \bibinfo {author} {\bibfnamefont {D.}~\bibnamefont {Dalacu}}, \bibinfo {author} {\bibfnamefont {V.}~\bibnamefont {Zwiller}},\ and\ \bibinfo {author} {\bibfnamefont {A.~W.}\ \bibnamefont {Elshaari}},\ }\bibfield  {journal} {\bibinfo  {journal} {arXiv}\ }\href {https://doi.org/10.48550/arXiv.2409.14964} {10.48550/arXiv.2409.14964} (\bibinfo {year}
  {2024}{\natexlab{a}}),\ \Eprint {https://arxiv.org/abs/2409.14964} {2409.14964} \BibitemShut {NoStop}%
\bibitem [{\citenamefont {Gao}\ \emph {et~al.}(2024{\natexlab{b}})\citenamefont {Gao}, \citenamefont {Krishna}, \citenamefont {Yeung}, \citenamefont {Yu}, \citenamefont {Gangopadhyay}, \citenamefont {Chan}, \citenamefont {Huang}, \citenamefont {Descamps}, \citenamefont {Reimer}, \citenamefont {Poole} \emph {et~al.}}]{gao2024demand}%
  \BibitemOpen
  \bibfield  {author} {\bibinfo {author} {\bibfnamefont {J.}~\bibnamefont {Gao}}, \bibinfo {author} {\bibfnamefont {G.}~\bibnamefont {Krishna}}, \bibinfo {author} {\bibfnamefont {E.}~\bibnamefont {Yeung}}, \bibinfo {author} {\bibfnamefont {L.}~\bibnamefont {Yu}}, \bibinfo {author} {\bibfnamefont {S.}~\bibnamefont {Gangopadhyay}}, \bibinfo {author} {\bibfnamefont {K.-S.}\ \bibnamefont {Chan}}, \bibinfo {author} {\bibfnamefont {C.-T.}\ \bibnamefont {Huang}}, \bibinfo {author} {\bibfnamefont {T.}~\bibnamefont {Descamps}}, \bibinfo {author} {\bibfnamefont {M.~E.}\ \bibnamefont {Reimer}}, \bibinfo {author} {\bibfnamefont {P.~J.}\ \bibnamefont {Poole}}, \emph {et~al.},\ }\href@noop {} {\bibfield  {journal} {\bibinfo  {journal} {arXiv preprint arXiv:2409.14964}\ } (\bibinfo {year} {2024}{\natexlab{b}})}\BibitemShut {NoStop}%
\end{thebibliography}%
\end{document}